\documentclass[12pt]{iopart}
\usepackage{iopams}  
\usepackage{amsmath,graphicx}

\begin{document}

\title[Critical Telomerase Activity]{Critical Telomerase Activity for Uncontrolled Cell Growth}

\author{Neil L Wesch,$^1$ Laura J Burlock$^1$\footnote{Present address:
Interdisciplinary School of Health Sciences, University of Ottawa, Ottawa ON K1N 6N5 Canada} and Robert J Gooding$^{1,2,3}$}
 
\address{$^1$Dept of Physics, Engineering Physics \& Astronomy,\\
$^2$Dept of Pathology \& Molecular Medicine,\\
$^3$Division of Cancer Biology \& Genetics, Cancer Research Institute,\\
Queen's University, Kingston ON K7L 3N6 Canada}
\ead{goodingr@queensu.ca}
\vspace{10pt}
\begin{indented}
\item[]May 2016
\end{indented}

\begin{abstract}
 
The lengths of the telomere regions of chromosomes in a population of cells are modelled using a chemical master equation
formalism, from which the evolution of the average number of cells of each telomere length is extracted.
In particular, the role of the telomere-elongating enzyme telomerase on these dynamics is investigated. We show that for biologically relevant rates of cell birth and death, one finds a critical rate, $R_{crit}$, of telomerase activity such that the total number of cells diverges. Further, $R_{crit}$ is similar in magnitude to the rates of mitosis and cell death. The possible relationship of this result to replicative immortality and its associated hallmark of cancer is discussed.

\end{abstract}

%
\vspace{2pc}
%
Published: \textit{Physical Biology {\bf 13} (2016) 046005}
%
\maketitle
%
%

\section{\label{sec:intro}Introduction}

A healthy mature individual undergoes cellular replication and division in a variety of tissue types and
blood cells. Associated with these events is the shortening of repeat units at the ends of each chromosome called telomeres. Telomeres were first recognized in the 1930's as special structures capping the ends of chromosomes, the cell's genetic material. Further research has identified telomeres to be important components in a variety of cellular processes, the most important being replication \cite{Blackburn 1991}.  Cellular replication involves one cell dividing itself into two identical copies, each containing a full copy of the genetic material. It was proposed \cite{harley_1992} that a generational clock may be associated with the progressive shortening of the telomeres. The decreasing lengths of telomeres have been found to be involved in aging, with shorter telomeres associated with advanced age. There is a complicated and not always consistent association
between telomere length and human diseases such as cancer \cite{chial_2015}.

The replication of the cell's linear chromosomes is a inherently destructive process, in that each replication causes a shortening of the chromosome through a phenomenon known as the end-replication problem  \cite{harley_1992,Olovnikov 1973}.  Cells are able to protect their (coding) genetic material by instead allowing telomeres to shorten \cite{Blackburn 1991}.  Further, in the absence of particular proteins and processes (discussed below) there exists a limited number of such divisions before which the cell reaches the so-called Hayflick limit \cite{Hayflick,Shay 2000}. Cells whose telomeres have been shortened to this critically short length are said to have reached replicative senescence. 

While almost all eukaryotic cells contain chromosomes and telomeres (read: not mature red blood cells), there are only some cell types that are actively dividing throughout adulthood. Other cells instead have lifetimes on the same order as the organism itself, and thus divide rarely or not at all. Leukocytes, or white blood cells, are among the cell types that regularly divide. Leukocytes encompass multiple different types of mono-nucleated blood cells including lymphocytes, \textit{viz.} B cells and T cells, as well as granulocytes. These cells are involved in the cellular component of immune response, and thus they experience large cellular turnover and are required to maintain large, dynamic  populations. (The function and behaviour and, in fact, the names of these cells change with repeated
sequences of replication.) The ease with which leukocytes can be extracted and analyzed, especially when compared to other rapidly dividing cell types, such as colon crypt cells, makes them the perfect subject for analyzing the Hayflick limit and telomere length dynamics.

All blood cells, white or red, originate as hematopoietic stem cells within the bone marrow. These stem cells are pluripotent, self-renewing cells capable of differentiating into unique types of hematopoietic progenitor cells, which then mature into peripheral blood cells \cite{Sidorov 2009}. The hematopoietic stem cells are considered to be ``immortal cells", and are believed to maintain their telomere length by means of the enzyme telomerase. The kinetics of B, NK and T cells, and other leukocytes has been studied extensively, and values have been obtained for their proliferation and the death rates.  Some estimated values are shown in Table 1.  

\begin{table}[h!]
\centering
\begin{tabular}{| l || c |  c |}
\hline
{\bf Cell Type} & {\bf Mitotic Rate (/week)} & {\bf Death Rate (/week)} \\ \hline\hline
Memory T-Cells  \cite{Borghans 2007}& 0.02 - 0.03   & 0.02 - 0.03   \\ \hline
Mature B-Cells  \cite{Cooperman 2004} & 0.5   & 0.48 \\ \hline
Neutrophiles \cite{Summers 2010}  & ~0.042  & ~0.042  \\ \hline
\end{tabular}
\caption{Estimated values from the literature for the mitotic and death rates for (memory) T cells, (memory) B cells and (neutrophiles) granulocytes.}
\end{table}

As cells divide and multiply the end-replication problem causes chromosomal shortening.  The level of attrition in leukocytes has been measured and most estimates of the number of base pairs lost per division range from 50 to 100 bp, with the highest estimate at 120 bp/division \cite{Weng 1995,vaziri 1993}.  This value is not uniform across species or even between individuals, with the exact figure changing greatly between individuals and different cells types.  Leukocytes  were found to be able to divide on average 23 times before reaching senescence \cite{pilyugin97}.  Similiar to the amount of base pairs lost per cell division, the Hayflick limit varies greatly among different cell types. For human fibroblasts between 14 and 29 divisions were observed, while for human embryos the limit was found to be between 35 and 63 divisions \cite{Hayflick}.

The role of the telomerase enzyme is to elongate telomeres, effectively mitigating the effects of telomere shortening. It is composed of an RNA structural protein complex termed TERC (hTERC in humans) and a catalytic protein complex known as TERT  (hTERT in humans). As with all proteins, the function is dependent upon the structure. Telomerase acts by positioning itself along the telomeric region of DNA with its intrinsic strand of RNA bases aligning with the $3'$ overhang of the chromosome. In humans the alignment function is carried out by hTERC. The intrinsic RNA component, hTERT in humans, is a single strand of RNA with base pairs which complement the nucleotides on the telomere. In the case of humans this corresponds to AAUCCC; this \textbf{motif} is repeated roughly three times (uracil replacing thymine in hTERT is not an issue, as both result in complimentary base pairing with adenine). Using standard biological syntax the RNA bases of hTERT would be expressed as $5'$-AAUCCCAAUCCCAAUCCC-$3'$. The labeling is due to the fact that telomerase elongates in the $5'$ to $3'$ direction. This leads to the addition of a $5'$-TTAGGGTTAGGGTTAGGG-$3'$ section being added to the overhang. DNA polymerase then elongates the complimentary strand, thereby elongating the double stranded section of the telomere. The positioning of DNA polymerase still requires an RNA primer, which is not replaced, thus the $3'$ overhang remains. The exact mechanism of this elongation was first elucidated in  1991  by Greider \cite{greider_1991}. 

There are many approaches possible to model the complex biology of telomere dynamics, and a recent in-depth review
of much of this work is included in a recently been published study \cite{hirt_2014}. Therefore, we refrain from providing
an exhaustive listing of previously published work. Of particular relevance to this paper are the models of Itzkovitz \emph{et al} \cite{itzkovitz_2008} and the revision of their model introduced by Cyrenne and one of us \cite{cyrenne_2015}. In particular,
these models examined the various properties of a population of cells that underwent replication and death that included
an index tracking the generational age of each cell. In this paper we extend these studies to include, in the simplest fashion
possible, the effect of the telomere-elongating protein telomerase on such populations.  The work of Hirt \emph{et al}  \cite{hirt_2014}, like our study, also incorporates telomerase into
the modelling. It provides a detailed theory predicting the distribution of telomere lengths and its dependence on telomerase.
However, unlike their work, our paper only seeks to provide the simplest possible mathematical representation of the role
of the variable activity of telomerase on these dynamics. That is, we follow the number of cells of each generational age at a given chronological time. Because of and in spite of the simplicity of our model, we are able to extract information regarding the dynamics that result if the expression  of the telomerase enzyme exceeds a critical value. Further, we show, subject to the limitations of  the coarse model we employ, that this critical value is not that different from the mitotic and cell death rates that are found in healthy cells. Other models have identified cell immortality, and in an elegant fashion include the role of stochasticity, such as that of Antal \emph{et al}  \cite{antal}. However, here we study a different mechanism of telomere
elongation, that is telomerase and not telomere sister chromatid exchange, and focus on how a divergence of cell population numbers are produced by the activity of telomerase.

\section{\label{sec:bioinform}Levels of Telomerase in Humans}

The literature includes ample qualitative discussion of telomerase levels in cancerous cells \cite{hiyama}, as well as the repeated statement that in almost healthy (non stem cell) tissues
the level of this enzyme is effectively zero  \cite{Vaziri_1998}. Modern sequencing technology allows for precise quantitative statements to be made, with minimal effort, so here we provide such information regarding the levels of this enzyme. 

The results shown here are in whole or part based upon data generated by the TCGA Research Network: http://cancergenome.nih.gov/. To be concrete, we have
examined RNASeqv2 data from matched normal/tumour samples for several cancers: bladder, ovarian, prostate and breast. The raw RNASeq data were analysed by the TCGA
consortium using MapSlice and RSEM algorithms \cite{mapsplice,rsem}, and our conclusions are based on Level 3 data. To be concrete, count data for 20,531 genes, annotated by RSEM and Entrez Gene numbers, are obtained. For all cancers that we studied, quanlitatively and often quantitatively similar results were obtained.

To place the numbers discussed below in context, note that all samples examined had average
gene expressions (counts) of between 900 and 1100. The maximum gene expression counts, typically for some collagens and housekeeping genes, was of the order of 10$^5$. The count levels for $hTERC$ and $hTERT$ were 1 to 3 orders of magnitude less than the 10$^3$ average expression level.

\begin{figure}[h!]
\centering
\includegraphics[height=7cm,width=9cm]{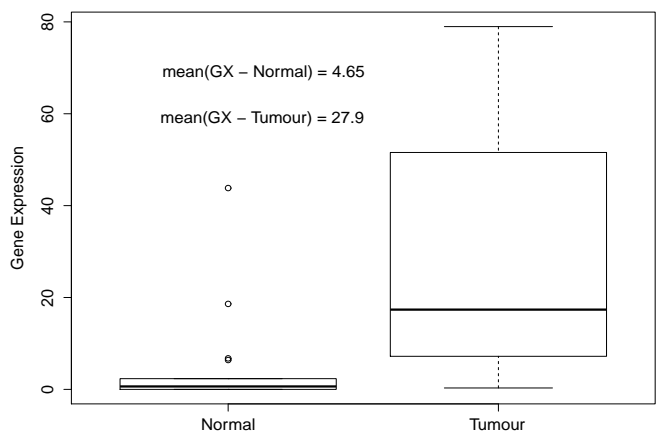}
\caption{\label{fig:BladderCa}A box-whisker plot of 18 matched normal/tumour samples for bladder cancer from TCGA. The heavy solid lines denote the medians, the boxes are the 25-75 quartile range,
the whiskers show 1.5 IQR in each direction, and the open circles are (outlier) data that lies outside of the range of the whiskers. The mean expression level (denoted by GX) in the tumours
is found to be six times that of the matched normal samples. Further, 7 out of the 18 normal samples have an (effective) expression level of zero.}
\end{figure}

We found in all samples that the counts for the telomerase RNA component, $hTERC$, were close to the lower detection level, usually of order one, and rarely as high as 10. There is
little difference in the $hTERC$ gene expression counts between normal and tumour samples. However, the telomerase reverse transcriptase $hTERT$ did show consistent changes between normal and tumour samples, with the  tumours having a mean expression level between 3 and 6 times that of the associated normal samples. (A comprehensive examination of all of this data can be found in Ref.~\cite{Gooding_2016}.) As one clear example, as of the time of the writing of this report there were 18 matched normal/tumour samples for bladder cancer at TCGA. In Fig.~\ref{fig:BladderCa} a box-whisker plot
of the expression data shows that one has a six-fold increase in the mean expression level for the tumours in comparison to the normal samples. Further, in 17 of the 18 samples the
expression level is higher in the tumour than the normal samples, whereas in the 18$^{th}$ sample the expression levels are effectively equal. Consistent with the literature, in 7 out of the 18 normal samples the expression level of $hTERT$ is effectively zero. 

Below we proceed with the understanding that the presence of  telomerase in samples is often associated with cancer, as supported by Fig.~\ref{fig:BladderCa}.  Experimental
identification of the activity rate of the elongation telomeres by telomerase is presently difficult to obtain, but could perhaps be provided $in~vitro$ by florescently tagging telomeres, and then
following their lengths using cytometry as a function of the concentration of $hTERT$.

\newpage

\section{\label{sec:mathm}Methods - Mathematical Modelling}

\subsection{Biochemical Reactions:}

For the models presented in this paper, the biochemical species of interest are \emph{cells} and \emph{telomerase proteins}. Within an approximation explained below, a cell is characterized by its generational age. This is the difference in the number of times a cell has undergone mitosis and replication compared with the number of times it has been acted upon by telomerase. A cell species will be denoted by $C$. The generational age is denoted by the subscript $i$. When a cell undergoes mitosis its generational age increases by one; that is $i \longrightarrow i+1$. When a cell is acted upon by telomerase its generational age decreases by one; that is $i \longrightarrow i-1$. A cell's generational age can run from $0$ (a newly differentiated cell) to $h$ (a senescent cell). Cell species of different generational age will be represented by $C_i$, and the respective population of cells of each generational age is denoted by $N_i$. These cellular populations are a function of time and will be termed the \emph{subpopulation functions}, $N_i(t)$, for $i=0,1,\dots,h-1,h$.

A telomerase protein species is represented by $T$. The telomerase population (number of proteins) is given by $N_T$. Then we denote the concentration of telomerase as $y(t)$ in units of number per volume, or concentration. 

The section of the telomeres that is lost during replication, and the lengths of telomeres that are regenerated
via telomerase, are not constant or equal.  Previous mathematical modelling by Cyrenne and one of us of a longitudinal study of baboons \cite{cyrenne_2015} found lengths between 120 and 600 base pairs could be lost during mitosis. Similar variations in the elongation by telomerase are expected - that is, there is a stochastic component to this biology. Therefore,
it is not \emph{per se} the number of mitotic divisions less the number of elongations that sets the Hayflick limit,
but rather a critical length of the telomeric region that induces replicative senescence. Here we simplify our
work without sacrificing the key result (regarding a critical rate of telomerase activation giving rise to a divergent number of cells) by taking the lengths of the telomeres lost during replication and the number of the TTAGGG
repeats added via the action of telomerase to be equal (in total length, in terms of base pairs), and the Hayflick limit to be determined by the net generational age of a cell. This approximation could change our result quantitatively, but not qualitatively, as we argue below.

A comment on terminology is appropriate before proceeding. \emph{Pristine cells} are those with generational age of zero, $i=0$ or $C_0$. \emph{Senescent cells} are those which have reached the Hayflick limit, $i=h$ or $C_h$. \emph{Permissive cells} include all those that are not either pristine or senescent, $i=1,\dots,h-1$ or $C_1,\dots,C_{h-1}$. We refer to them as permissive because these cells \emph{permit} mitosis and replication to occur, and consequently permit the total number of cells in the population to increase, the crucial element needed to make sense of our results.

We now describe and model the relevant biochemical reactions that occur in a population of replicating cells. First, mitotic replication is taken to occur at a rate $M$ of replicative events per cell per unit time. This process results in one cell of generation $i$, $C_i$, producing two cells of generation $i+1$, $2C_{i+1}$. This is shown in  Eq.~(\ref{subeq:mitosis_reaction}), and all
subsequently described processes, first in terms of the chemical reaction, and then for the changes induced in (instantaneous)
populations, $\{N_i\}_{i=0}^h$.
\begin{eqnarray}
\label{subeq:mitosis_reaction}
C_i  & \xrightarrow{~~ M ~~  } & ~~~~2~C_{i+1} ~~~~~~~~~~\textrm{for}~~i=0,\dots,h-1\\
(N_i,N_{i+1}) & \xrightarrow{~~  M ~~ } & ~~~~(N_i-1, N_{i+1}+2)\nonumber
\end{eqnarray}
As is made clear above, only $i\not=h$ cells can undergo mitosis, and the inability of a cell whose chromosomes have
reached the Hayflick limit to subsequently divide is often referred to as replicative senescence \cite{Hayflick}.

Acute death removes a cell from the derived population pool. This occurs at a rate $D$ of mortality events per cell per unit time. Therefore a cell of generation $i$, $C_i$, is removed from the system, as shown in Eq.~(\ref{subeq:death_reaction}).  
\begin{eqnarray}
\label{subeq:death_reaction}
C_i \ & \xrightarrow{~~ D ~~  } &~~~~\emptyset ~~~~~~~~~~\textrm{for}~~i=0,\dots,h\\
(N_i) \ & \xrightarrow{~~ D ~~  } &~~~~  (N_i-1)\nonumber
\end{eqnarray}
Note that in all reactions the null set $\emptyset$ denotes the lack of presence in the population pool of cells. Cells of any generation are susceptible to acute death, and therefore these expressions are valid for $i=0,\dots,h$.

The population pool is ``filled" via the differentiation of stem cells. The rate at which cells leave the stem cell population pool and enter the population pool is denoted by $\alpha$, and is in number of cells per unit time. Collectively this results in the production of new $C_0$ cells and an increase in $N_0$ as shown below.
\begin{eqnarray}
\label{subeq:differentiation_reaction}
\emptyset  &  \xrightarrow{~~ \alpha ~~  } &~~~~ C_0 \\
(N_0) \ & \xrightarrow{~~ \alpha ~~  } &~~~~ (N_0+1)\nonumber
\end{eqnarray}

Through the previously discussed mechanisms, telomerase acts to ``rejuvenate" a cell at a rate $R$ per cell per unit time. The specific details as to what we choose to take as $R$ will be discussed in more detail below. The net result is the onversion of a cell of generation $i$, $C_i$, to a cell of generation $i-1$, $C_{i-1}$ (it is in this sense that telomerase rejuvenates a cell - it makes it younger).  The effect of telomerase as we model it in this paper is contained in Eq.~(\ref{subeq:telomerase_reaction}).  
\begin{eqnarray}
\label{subeq:telomerase_reaction}
T + C_i  &  \xrightarrow{~~R ~~  } &~~~~ T + C_{i-1} ~~~~~~~~~~\textrm{for}~~i=1,\dots,h\\
(N_i, N_{i-1})  & \xrightarrow{~~R ~~  } &~~~~  (N_i-1, N_{i-1}+1)\nonumber
\end{eqnarray}
Only $i\not=0$ cells can be acted upon by telomerase. A telomerase protein, $T$, is essential to the reaction, but is not consumed in the process. (A model in which $R$
is the only non-zero rate is presented in the appendix.)

\subsection{Chemical Master Equation:}

We consider the system to be composed of these reactions with these \textit{average} reaction rates. Also,
for now we take the concentration of telomerase to be fixed, but return to the lifting of this simplification later in
this paper. The probability
of finding the system in a state having $N_0,~N_1,~\dots,~N_h$ cells with these number of net mitotic divisions is
denoted by $P(N_0,~N_1,~\dots,~N_h,t)\equiv P(N_i,t)$. Following this notation, the chemical master equation for such a collection of cells and telomerase can be written as (here we suppress the time dependence of the probability function)
\begin{equation}
\label{eqn:chemical_master_equation_general}
\begin{split}
\frac{ \partial P(\{N_i\})}{\partial t} \ = \ & \ \alpha~ \left[ P(\{N_0+1\}) - P(\{N_i\}) \right] \\
& + D~ \sum_{i=0}^{h} \left[ (N_i + 1)P(\{N_i+1\}) - N_iP(\{N_i\}) \right] \\
& +  R~ \sum_{i=1}^{h} \left[ (N_i + 1)P(\{N_i-1\}) - N_iP(\{N_i\}) \right] \\
& + M~ \sum_{i=0}^{h-1} \left[ (N_i+1)P(\{N_i+1,N_{i+1}-2\}) - N_iP(\{N_i\}) \right] \\
& + A~ \left[ (N_h-1)P(\{N_h-1\}) - N_hP(\{N_h\}) \right]
\end{split}
\end{equation}
Note the differing summation limits.

\subsection{Deterministic Equations for the Average Subpopulation Functions:}

If all reaction rates are taken to be constant, then one may derive the equations of motion of the average
subpopulation functions. We denote averages over an ensemble of populations of cells by $\{{\mathcal N}_i\}_{i=0}^h$. 
Then one finds the following equations of motion for these average subpopulation functions to be simple coupled
linear ODES, namely
\begin{eqnarray}
\label{subeqn:STM_dynamical_equations}
\dot{\mathcal N}_0(t) & = &  \alpha - (M+D)~{\mathcal N}_0(t) + R~{\mathcal N}_1(t) \nonumber \\
\dot{\mathcal N}_i(t) & = &  2M~{\mathcal N}_{i-1}(t) - (M+D+R)~{\mathcal N}_i(t) + R~{\mathcal N}_{i+1}(t) ~~~~ i=1,\dots,h-1 \nonumber \\
\dot{\mathcal N}_h(t) & = &  2M~{\mathcal N}_{h-1}(t) - (D+A+R)~{\mathcal N}_h(t)  
\end{eqnarray}

One may examine many aspects of the subpopulation functions, including the average length of the telomeres for the chromosomes
in each cell, and we discuss some of these results below. However, the main focus of this paper is in the (average) total number of
cells in the population, which we denote by $n(t)$ and is defined by
\begin{equation}
n(t)~=~\sum_{i=0}^h~{\mathcal N}_i(t)
\end{equation}
Conditions under which this total population size becomes very large, which physically could correspond to a rapidly growing tumour and which mathematically we will associate with a diverging population size, is related to the progression of cancer\cite{HOC_1,HOC_2}. Below we identify conditions under which our simple model can lead to such divergences.

\subsection{Limiting Behaviour -- Previously Published Models:}

One of the focuses of the results given below  is concerned with the long-time limit of the total (average) population, $n(t)$. To place these new results properly in perspective, here we summarize this limiting behaviour using previously published work.

Itzkovitz \textit{et al} \cite{itzkovitz_2008} examined a model with no Haylflick limit ($h\rightarrow\infty$) with the constraint that
the total population is fixed at its $t=0$ value. For the initial condition corresponding to only $i=0$ cells being present at $t=0$,
namely $n(0)=N_0(0)$, this situation is guaranteed if $\alpha=(M+D)n(0)$. One may generalize their model to allow
for non-constant populations, and for the specific case of an empty initial state ($n(t=0)=0$), iteratively solving the associated ODEs one finds
\begin{equation}
n_{h\rightarrow\infty}(t)~=~\frac{\alpha}{M+D}~\sum_{j=0}^\infty~\Big(\frac{2}{1~+~D/M}\Big)^j
\end{equation}
Therefore, for the situation of interest below, namely $M>D$, this sum diverges to $\infty$. We will relate this limit to our work below.

If one modifies the model of Ref.~\cite{itzkovitz_2008} to include the Hayflick limit then one may follow the work of Cyrenne and one of the present authors in Ref.~\cite{cyrenne_2015} to obtain
\begin{equation}
\label{eq:cyrenne_noft}
n_{h<\infty}(t)~=~\frac{\alpha}{M+D}~\sum_{j=0}^h~\Big(\frac{2}{1~+~D/M}\Big)^j~+~\frac{\alpha}{D}~ \frac{2^hM^{h-1}}{(M~+~D)^{h+1}}
\end{equation}
Clearly, this latter result makes clear that the steady state average total population will always remain finite. One may connect
the formalisms of Refs.~\cite{itzkovitz_2008} and \cite{cyrenne_2015} by noting that in enforcing the existence of the Hayflick limit
the population size will remain finite (as  it does in  \cite{itzkovitz_2008}) but will approach its steady state value from some arbitrarily defined initial $t=0$ distribution.

\subsection{Main result of study:}

The noteworthy result of the present paper is simply that Eq.~(\ref{eq:cyrenne_noft}) will be modified such that $n_{h<\infty}(t\rightarrow\infty)\rightarrow\infty$ due to activity of telomerase {\it when} the activity of this protein exceeds a critical value. The structure of the equations of motion for the average populations of each generational age does not make this obvious. That is, the action of telomerase is number conserving - the result of its action is simply to change what we are calling the generational age of a cell from $i$ to $i-1$. If no new cells are created because of $R$, why is the population size so drastically affected? We provide one explanation for this behaviour in the following section.

\newpage

\section{Results - Uncontrolled Cell Growth and $R_{crit}$}

\subsection{Dynamics of total population $n(t)$:}
\label{subsec:dynnoft}

We now turn to the main result of our paper. We show that for a model that includes the Hayflick limit ($h$ finite) if $M>D$ then
a critical value of $R$ exists, which we denote as $R_{crit}$, such that $n(t\rightarrow\infty)\rightarrow\infty$.

\begin{figure}[h]
\centering
\includegraphics[height=8cm,width=9cm]{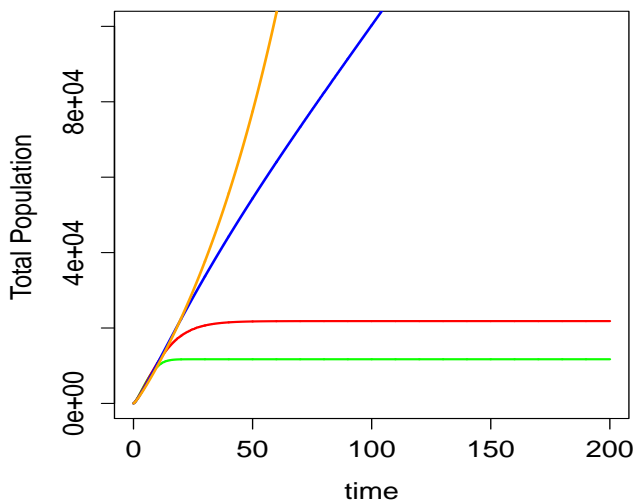}
\caption{\label{fig:STM_n_mitosis_dominant}Plots of the total population function $n(t)$ for $h=25$ for mitotic and death rates appropriate for mature B white blood
cells, and increasing values of $R$. $R=0$ is shown in green (lower curve), $R=0.4$ is shown in red (2nd lowest curve), $R=0.8$ is shown in blue (2nd highest curve), and $R=1.2$ is shown in orange (highest curve). For these we have set $\alpha=400, M=0.50,  D=0.48$ and $A=0$, and we find that $R_{crit}=0.785$.}
\end{figure}

Example plots of this situation are shown in Fig.~\ref{fig:STM_n_mitosis_dominant} for the initial state at $t=0$ being that no cells are present in the derived population pool. Note that irrespective of the choices
of $\alpha,~M,~D$, when $h$ is finite and $R=0$ then $n(t\rightarrow\infty)$ always remains finite - see Eq.~(\ref{eq:cyrenne_noft}). However, as the curves in Fig.~\ref{fig:STM_n_mitosis_dominant} make clear, only
when $R$ exceeds some critical value (here between 0.5 and 1) does the (average) total number of cells diverge. We label this critical value of $R$ as $R_{crit}$.

The identification of $R_{crit}$ is one of the key results of this paper. The average
total population of all cell types can be shown to satisfy the following equation:
\begin{equation}
\label{eq:ndot}
\dot n(t)~=~\alpha ~+~(M-D) n(t) ~-~M~\mathcal N_h (t)
\end{equation}
However, the solutions for the subpopulations such as $\mathcal N_h (t)$ via, \textit{e.g.} Laplace transforms, are cumbersome. Therefore, in order to most easily identify $R_{crit}$ the form of the analytic matrix solutions to these deterministic ODEs will be presented. The  equations of motion of the average populations givens
in Eq.~(\ref{subeqn:STM_dynamical_equations}) can be expressed in an equivalent form using matrices and vectors:
\begin{equation}
\label{eqn:STM_matrix_system}
\dot{\underline{\mathcal N}} ~ = \underline{\underline{Z}} \cdotp \underline{\mathcal N} + \underline{\alpha}
\end{equation}
where 
\begin{equation}
\underline{\mathcal N}(t) = 
		\begin{bmatrix}
		\mathcal N_0(t) \\
		\mathcal N_1(t) \\
		\mathcal N_2(t) \\
		\dots \\
		\mathcal N_h(t)
		\end{bmatrix} ~~~~~~~~~~~~~~~~~
\label{subeqn:STM_matrix_alpha}
\underline{\alpha} = 
		\begin{bmatrix}
		\alpha \\
		0 \\
		0 \\
		\dots \\
		0
		\end{bmatrix}
\end{equation}
and the (square non-symmetric) coefficient matrix is denoted by $\underline{\underline{Z}}$, and can be written down
for a given $h$ using Eq.~(\ref{subeqn:STM_dynamical_equations}). To allow for the application of a helpful theorem, it is informative to write down the matrix for a given $h$, 
so for $h=3$ this matrix is shown below. (Note that the tridiagonal form of this $(h+1)\times(h+1)$ matrix is preserved in rows 2 to $h$.)
\begin{equation}
 \underline{\underline{Z}} = 
		\begin{bmatrix}
		-M-D & R & 0 & 0 \\
		2M & -M-D-R & R & 0 \\
		0 & 2M & -M-D-R & R \\
		0 & 0 & 2M & -D-R
\end{bmatrix}
\end{equation}

Equation~(\ref{eqn:STM_matrix_system})  is solved using conventional methods for ordinary differential equations - one must solve both the homogeneous and inhomogeneous equations. The solution to the homogenous equation includes an expansion of eigenvectors and eigenvalues, and the eigenvalue solutions for the $\underline{\underline{Z}}$ matrix will be denoted by
\begin{equation}
 \underline{\underline{Z}}~\underline{v}_j~=~\lambda_j~\underline{v}_j
\end{equation}
The complementary particular solution is a vector of constants, which we denote as $\underline{k}$, and is given by  \begin{equation}
\label{eqn:alpha_particular}
\underline{k}
\ = \
- ~\underline{\underline{Z}}^{-1} \cdotp \underline{\alpha}
\end{equation}
Then, letting the $c_j$ be expansion coefficients determined by initial conditions, the solution to Eq.~(\ref{eqn:STM_matrix_system}) can be written as
\begin{equation}
\label{eqn:STM_matrix_subpopulation_general}
\underline{\mathcal N}(t) = \underline{k} + \sum_{j=1}^{h+1} ~c_j ~\underline{v}_j ~e^{\lambda_j t}
\end{equation}

The divergent total number of cells seen in Fig.~\ref{fig:STM_n_mitosis_dominant} is therefore seen to be associated with one of the eigenvalues of $\underline{\underline{Z}}$ becoming positive. In particular, this mathematical formulation makes clear that this diverging $n(t)$ is independent of the initial conditions for the $\{\mathcal N(t=0)\}_{j=0}^h$ (unless one is considering the highly
unlikely initial condition such that the coefficient $c_j$ associated with the eigenvector with the positive eigenvalue is
equal to zero). Further, the divergence is not dependent on the existence of a pool of repopulating stem cells, since
$\underline{\underline{Z}}$ does not depend on $\alpha$.

\begin{figure}[t]
\begin{center}
\includegraphics[scale=0.55]{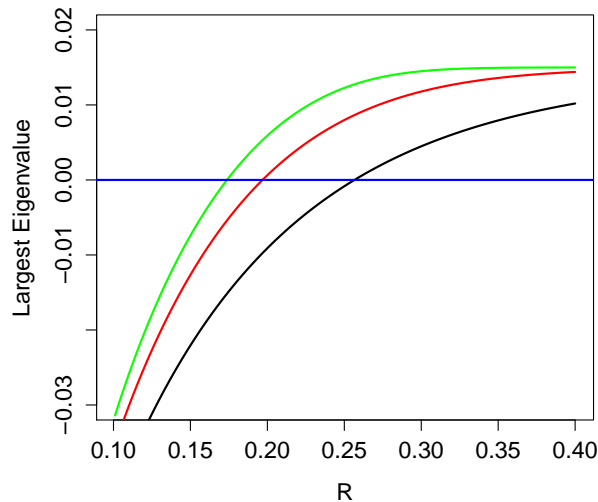}
\caption{\label{fig:largesteigs}A plot of the largest eigenvalues of the coefficient matrix  $\underline{\underline{Z}}$
as a function of $R$ for $h=5$ (lower black curve), $h=10$ (middle red curve) and $h=20$ (upper green curve) for
$M=0.14$ and $D=0.125$ (and $A=0$). The horizontal (blue) curve is a guide to the eye that indicates where the eigenvalues first become positive. For the value of $h=23$ one finds $R_{crit}=0.173$.}
\end{center}
\end{figure}

One may make progress analytically on the eigenvalue problem using the Gerschgorin Disk Theorem \cite{disktheorem}. For the 4$\times$4~$h=3$ matrix shown above, the real value of three of the eigenvalues is bounded above by $M-D$ (for the other eigenvalue the upper bound is zero). Therefore, we immediately have the result that only for $M>D$ can one obtain an eigenvalue with a positive real part. It is straightforward to generalize the application of this theorem to any $h>1$. 

Example results for the eigenvalue which becomes positive, leading to the divergent total population, are shown in Fig.~\ref{fig:largesteigs}. The variation of the eigenvalue
is shown as a function of $R$ for increasing values of $h$.
Here we take values based on a longitudinal study for baboon T cells \cite{Baerlocher 2007} completed previously ~\cite{cyrenne_2014}, namely $h=23,~M=0.14,$\break $D=0.125$. This
plot makes clear that as $h$ increases towards its known value ($h=20\sim25$), the critical value of $R$ at which
the eigenvalue first becomes positive decreases. For $h=23$ for these values we find $R_{crit}=0.173$. Similar
results are found when values for mature B cells are used, namely for $h\approx 25,M=0.5,~D=0.48$ we find 
$R_{crit}=0.785$. That is, when the mitotic replication rate $M$ and the cell death rate $D$ are similar, then $R_{crit}$ is of the same order of magnitude, but larger than either $M$ or $D$. We discuss these magnitudes below.

We have undertaken a numerical study (hopefully exhaustive ) examining the eigenvalues of $\underline{\underline{Z}}$ for biologically
relevant $h$, and rate constants $M$ and $D$ (and $A$). First, confirming the above analytical result, the diverging population occurs only for $M>D$. 
Second, if one augments the model to include so-called apoptotic death, denoted by $A$, meaning
the increased (net) death rate of senescent cells due to p53 triggered apoptosis, $R_{crit}$ is increased, but only by a small amount for biologically relevant $h$ (for a variety of cell types, not simply leukocytes, for which most experimental data is available). For example, for $h=23,~M=0.14,~D=0.125$, the parameters fit to the longitudinal baboon leukocyte data,
$R_{crit}$ increases from 0.173 for $A=0$ to 0.174 when $A=D$ is used.
Third, although the differential equation for $\dot n(t)$ may be written in a form that does not include $R$, such as Eq.~(\ref{eq:ndot}), the subpopulation solutions (such as $\mathcal N_h(t)$) do depend on $R$. Therefore, although $R$ itself conserves the number of cells (see the Appendix and Eq.~(\ref{subeq:telomerase_reaction})), its role is to transform cells that have become senescent, and therefore unable to replicate through mitosis, to subsequently become permissive thereby permitting a non-dividing senescent cell to again divide. This serves as the basis for the model that we present in the next subsection.  

\subsection{Concentration-Dependent Telomerase Activity:}
\label{subsec:Rdepends}

Our bioinformatics summary of the gene expression levels of the $hTERT$ transcript in matched normal/tumour samples, given earlier, suggests that it is a reasonable assumption to
say that only when a tumour is present does one find appreciable levels of telomerase. For example, during the process of oncogenesis the gene which is translated to produce telomerase is upregulated \cite{harley_1992,shay_1997}. Therefore, our previous model, which took the activity level of telomerase as $R$ to be a constant
was an assumption that we should investigate.

Instead, consider the following. Telomerase proteins, $T$, are produced at a rate $\beta$ and degrade at a rate $\Gamma$. Production of telomerase is accomplished through regular protein translation, and degradation of telomerase occurs through simple  denaturing mechanisms.
\begin{eqnarray}
\label{subeq:telomerase_production}
\emptyset  & \xrightarrow{~~ \beta ~~  } &~~~~ T \\
T  & \xrightarrow{~~ \Gamma~~  } &~~~~\emptyset 
\end{eqnarray}

The results in \S \ref{subsec:dynnoft} given above are those produced in our model when the concentration of telomerase
is a constant.However, the rate $R$ of elongating chromosomes is dependent on $y(t)$, the concentration of telomerase. Here we propose the simplest modelling of this concentration dependence that seems likely. 

The telomerase concentration function, $y(t)$, can be described by the following first-order ordinary differential equation
\begin{equation}
\label{eqn:CM_concentration_ODE}
\dot y(t) \ = \ \beta \Theta(t-t_o) - \Gamma y(t)
\end{equation}
$\beta$ represents the rate of production of the enzyme in number of proteins per unit time, whereas $\Gamma$ represents the rate of degradation of the enzyme in number of proteins per unit time. We have taken the production and decay rates to be concentration independent (although generalizations of this approach produce no important modifications
to our results).   We introduce a value we have termed \emph{the activation time} and denote it as $t_o$, and take
it to represent the time when telomerase production begins, such as some time corresponding to the appearance of a neoplasia.
Defining $y(0) \equiv y_0$ and physically assuming $t_o > 0$ gives 
\begin{equation}
\label{subeqn:CM_concentration_function_simplified}
y(t) \ =  \ y_0 e^{-\Gamma t} + \frac{\beta}{\Gamma} \left[1~-~ e^{-\Gamma \left( t - t_o \right) }   \right] \Theta \left( t - t_o \right) \\\end{equation}
Clearly, after some time $t\gg t_o$ the concentration of telomerase is fixed to be\\ $y(t\rightarrow\infty)=\frac{\beta}{\Gamma} $.

Now one must include the effect of this time dependence in the activity of telomerase. Therefore
we now introduce a new form of the telomerase rate \emph{functional} which implements an activator Hill function,  Eq.~(\ref{eqn:ACM_functional}) - this is the standard form for modelling enzymatic activity which increases non-linearly from a baseline. 
\begin{equation}
\label{eqn:ACM_functional}
R[y(t)] \equiv \frac{ r \ y(t)^H }{ k + y(t)^H }
\end{equation} 

Now consider the following scenario. A healthy post-puberty individual has a sufficiently small concentration of
telomerase that we can approximate $R\approx0$. Therefore, a system of cells (for some given tissue with
a stem cell source of new $j=0$ cells) reaches steady values of ${\mathcal N}_j(t\rightarrow\infty)\}_{j=0}^h$. 
Then crisis occurs at $t=t_o$, and quickly the rate $R$ increases and stabilizes at some new value. If $R[y(t\rightarrow\infty)]>R_{crit}$, one finds uncontrolled growth for
the affected tissue. That is, in comparison to the results
shown at the beginning of this section no qualitative change occurs, only the onset of the uncontrolled growth is shifted in time
to $t>t_o$.


\section{Conclusions and Discussion}

Using the simplest model possible, we have incorporated the effect of telomerase being present in a cell's nucleus, thereby allowing it to elongate the telomeres of the
cell's chromosomes. We have made many approximations in our treatment of this system, but none of them should influence our main result, namely that
if the activity of telomerase exceeds some critical value then the total number of cells will be found to diverge as $t\rightarrow\infty$. In the parlance of
the oft-quoted ``Hallmarks of Cancer" \cite{HOC_1,HOC_2}, these cells have obtained replicative immortality.

The critical value, $R_{crit}$, that we find is similar in magnitude to the average rate of cell division (mitosis) $M$, which is often similar but larger than the average rate of cell death $D$. The work of Ref.~\cite{hirt_2014} found a similar result, namely that the lengths of telomeres are elongated by telomerase, but only for unphysical levels of telomerase. Our critical value seems to be physical in that the size of $M$ is an effective rate, one that incorporates many biological processes,  and represents that all of hte breakpoints are traversed, which depend on many protein/enzyme levels, cell division occurs. If the population of cells with different telomere lengths is going to have an exponentially increasing number of total cells, bypassing the steady-state levels found in replicative senescence (for our parameter values $\mathcal{N}_h(t\rightarrow\infty)$ dominates the total number of cells in steady state), then telomere-elongation events must at least cancel out the byproducts of cell division, so finding $R_{crit}\sim M$ is indeed the expected result. 

Our model, the equations of motion for the average number of cells of each net generational age, is deterministic. However, no qualitative changes occur when one uses a stochastic simulation of the chemical master equation presented in this work (we used the full Gillespie simulation algorithm \cite{Gill1,Gill2}). Details of such work, in combination with other biological processes important in the initiation of the cell cycle, will be presented elsewhere.

The biological implications of these mathematical results must be considered carefully. No cell lasts forever in a living organism, so $n(t\rightarrow\infty)\rightarrow\infty$ is
not a directly useful limit. Instead, suppose that a healthy population of cells finds that a few cells obtain a 
replicative advantage, such as when a neoplasia forms. If the enzyme telomerase is activated in that population, then the advantaged cell won't necessarily be limited by the Hayflick limit, 
since the role of telomerase is, in part, to make senescent cells become permissive of mitosis. That is, the results from the present model should be considered in terms of 
an increasing number of selectively advantaged tumour cells without bound. If such cells
grow to some large size (think of the grapefruit size tumour often found in women with ovarian cancer) the consequences of some of these cells undergoing an
epithelial-mesenchymal transition thereby possibly leading to distant metastatic cells is the life-determining upper bound in time imposed by the cancer.

\section[]{Acknowledgments}
We thank an anonymous referee for alerting us to Ref. \cite{disktheorem}, to David Berman and Palak Patel regarding experiments to measure
the activity of telomerase, and to Troy Day for helpful comments. This work was supported in part by Queen's University.

\newpage
\appendix
\section[]{}

To further investigate the dynamics of our model, a limiting case is considered. That is, we now examine the same deterministic equations which were presented in \S\ref{subsec:dynnoft}, but for the case in which $R$ is the only non-zero rate constant. That is, in terms of applicable biological processes only telomerase acts on cells in the derived population pool. That is, we consider the model defined by
\begin{equation}
\label{eqn:SatTM_reaction}
(N_i , N_{i-1}) \ \overset{R}\longrightarrow \ (N_i -1 , N_{i-1}+1) \hspace{12mm} i=1,\dots,h
\end{equation}
(Note that for completeness this limit is examined. It does not correspond to a physical limit.) This is equivalent to setting 
$\alpha=M=D=0$ and  Eq.~(\ref{subeqn:STM_dynamical_equations}) reduces to   
\begin{eqnarray}
\label{subeqn:SatTM_dynamical_equations}
\frac{d \mathcal N_0(t)}{d t}  &= &  R~\mathcal N_1(t) \nonumber\\
\frac{d \mathcal N_i(t)}{d t}  &= &  - R~ \mathcal N_i(t) + R~ \mathcal N_{i+1}(t)  ~~~~~~~~~~i=1,\dots,h-1 \nonumber\\
\frac{d \mathcal N_h(t)}{d t}  & = &  -R~ \mathcal N_h(t)
\end{eqnarray}
These deterministic equations can be solved analytically by reverse iteration once a value for $h$ has been specified. To proceed, a closed form expression for $\mathcal N_h(t)$ can be calculated. This expression can then be used to solve for $\mathcal N_{h-1}(t)$. Repeating this process allows for all subpopulation functions to be obtained, ending with $\mathcal N_0(t)$.

We begin with a constant initial condition - it partitions the number of total cells evenly among all the available generations at $t=0$ and is given by $\mathcal N_i(0) = n_0 / (h+1)$ for all $i$.  When solving the resulting equations with this initial condition, each of these expressions is a geometric series, and this set of equations is that they hold true for any value of $h$.
\begin{eqnarray}
 \mathcal N_h(t) \ &= & \ n_0 e^{-Rt} \nonumber \\
 \mathcal N_i(t) \ &= & \ n_0 \left[ \sum_{j=0}^{h-i} \frac{(Rt)^j}{j!} \right] e^{-Rt} ~~~~~~~~~~~~~~ i=1,\dots,h-1 \nonumber  \\
 \mathcal N_0(t) \ &= & \ n_0 \left[ (h+1) - \sum_{j=0}^{h} (h-j) \frac{(Rt)^j}{j!} \right] e^{-Rt}
\end{eqnarray}
These equations are consistent with
\begin{equation}
n(t)~=~(h+1)~n_0
\end{equation}
for all times.

\end{document}